\documentclass[5p,twocolumn,times,number]{elsarticle}
\usepackage{amsmath}
\usepackage{amssymb}
\usepackage{graphics}
\usepackage{graphicx}
\usepackage{epsfig}
\usepackage{dcolumn}
\usepackage{bm}
\usepackage{lineno}
\begin{document}
	\begin{frontmatter}
		
		
		\title{Visual investigation of possible degradation in GEM foil under test}
		
		\author[]{S.~Chatterjee\corref{cor}}
		\cortext[cor]{Corresponding author}
		\ead{sayakchatterjee@jcbose.ac.in, sayakchatterjee896@gmail.com}		
		\author[]{A.~Sen}
		\author[]{S.~Das}
		\author[]{S.~Biswas}
		\address[]{Department of Physics, Bose Institute, EN-80, Sector V, Kolkata-700091, India}

		\begin{abstract}			
			
			Visual investigation of a Single Mask (SM) Gas Electron Multiplier (GEM) foil, showing small resistance~($\sim$~40~k$\Omega$), is performed manually using an optical microscope. The GEM foil is scanned and the different imperfections in the foil are identified. Different techniques are used to clean the GEM foil to remove the short paths created between the GEM electrodes. The details of the method used for cleaning the GEM foil and the result of leakage current measurement after the cleaning of the foil are discussed in this article.			
		\end{abstract}
		
	\end{frontmatter}
	
	\section{Introduction}\label{intro}
	\vspace*{-0.205cm}
	The Gas Electron Multiplier~(GEM) detector is widely used as a tracking device in High Energy Physics~(HEP) experiments due to its good position resolution~($\sim$~70~$\mu$m) and high rate~($\sim$~1~MHz/mm$^2$) handling capability~\cite{ref1}.  Depending on the photo-lithographic technique used, the GEM foils can be classified as either Double Mask~(DM) or Single Mask~(SM) foils~\cite{ref1_1}. Long-term stability is one of the important criteria for any detector used in HEP experiments~\cite{ref4,ref5}. 
	During the long-term test with a SM triple GEM chamber, it is observed that the detector suddenly stopped giving the signal. To understand the problem, the triple GEM chamber prototype is disassembled and the individual foil resistance is measured. It is found that the resistance of the 3$^{rd}$ GEM foil is $\sim$~40~k$\Omega$ which indicated that there are some short paths created between the top and bottom electrodes of the foil. The short-circuited path might be created due to the accumulation of impurities inside the GEM holes or due to the degradation of the foil itself. Visual investigation of the foil is performed manually using an optical microscope~(Nikon eclipse Ni) having magnification factors of 20X and 40X. After the optical scanning, the foil is cleaned using millipore water and with an ultrasonic~($\sim$~20~kHz) bath~\cite{ref6}. After cleaning, the foil resistance and the leakage current of the foil are measured. The optical inspection of the foil and the cleaning methodologies are discussed in sec.~\ref{section2} and sec.~\ref{section3} respectively. The setup and results of the leakage current measurement of the foil are discussed in sec.~\ref{section4}. 
	
	\vspace*{-0.3cm}
	\section{Optical inspection of the GEM foil}\label{section2}
	\vspace*{-0.1cm}
	The damaged GEM foil is investigated visually using an optical microscope with different magnification settings.
	\begin{figure}[htb!]
		\begin{center}
			\includegraphics[scale=0.059]{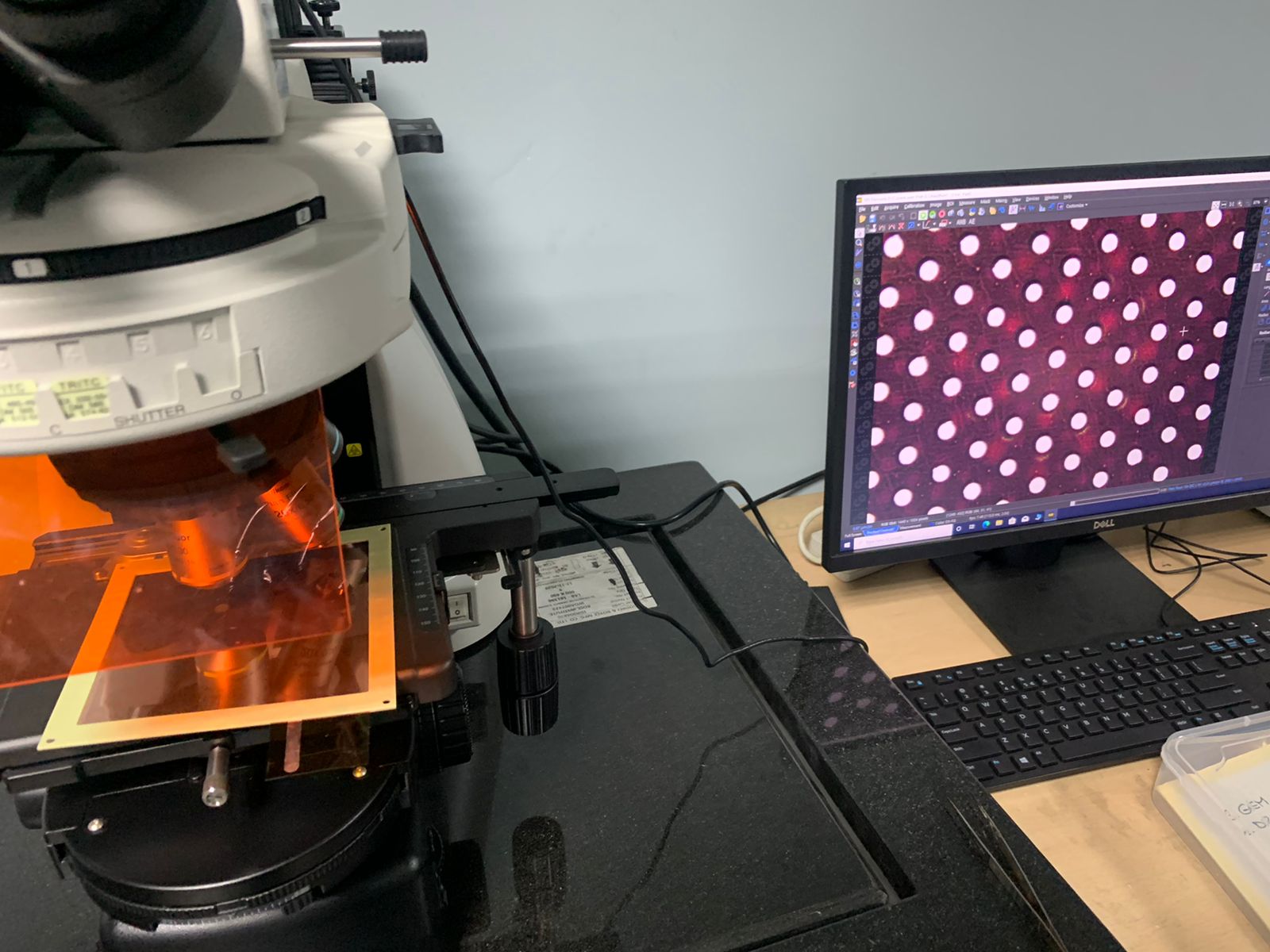}
			\includegraphics[scale=0.07]{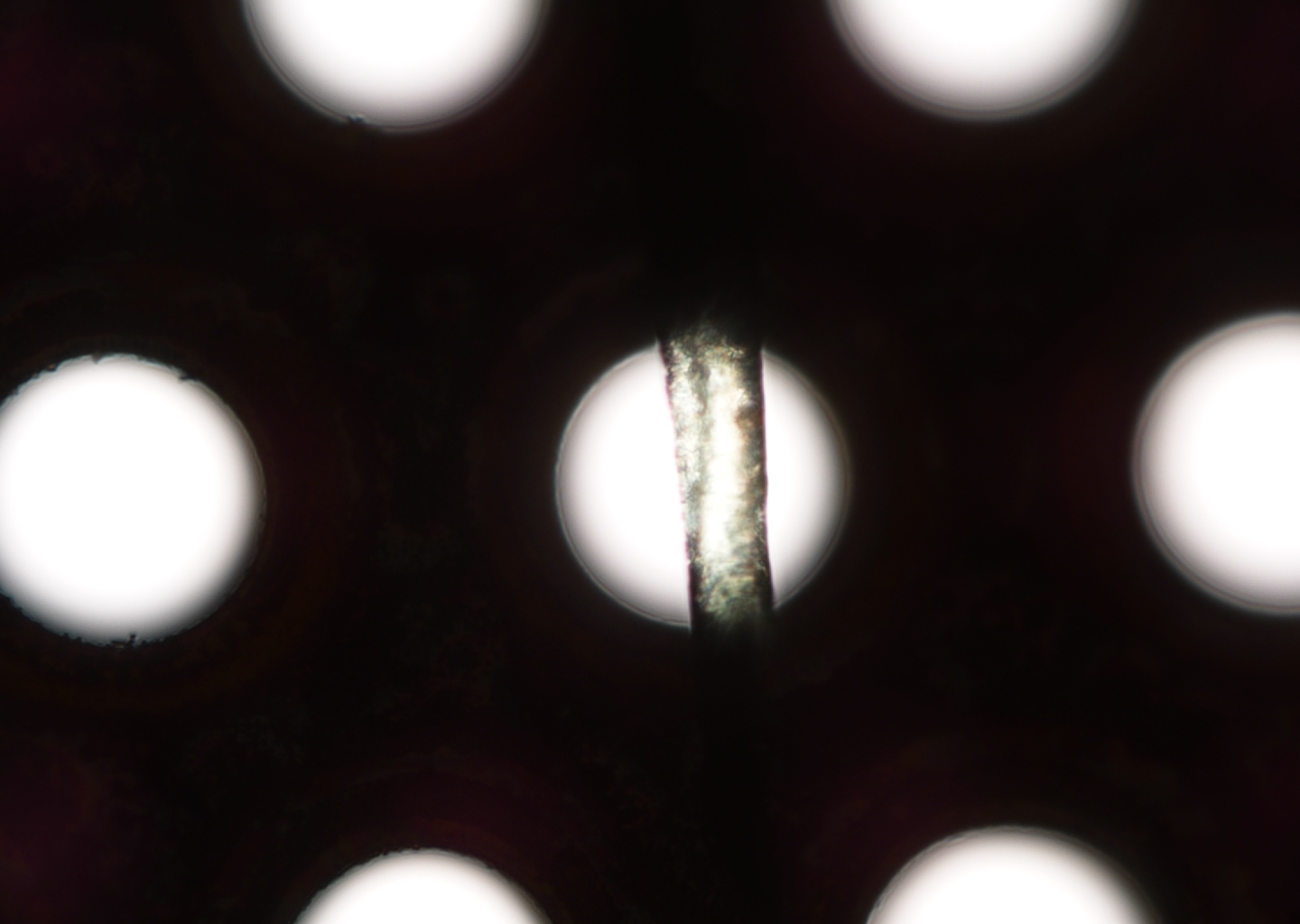}
			\includegraphics[scale=0.07]{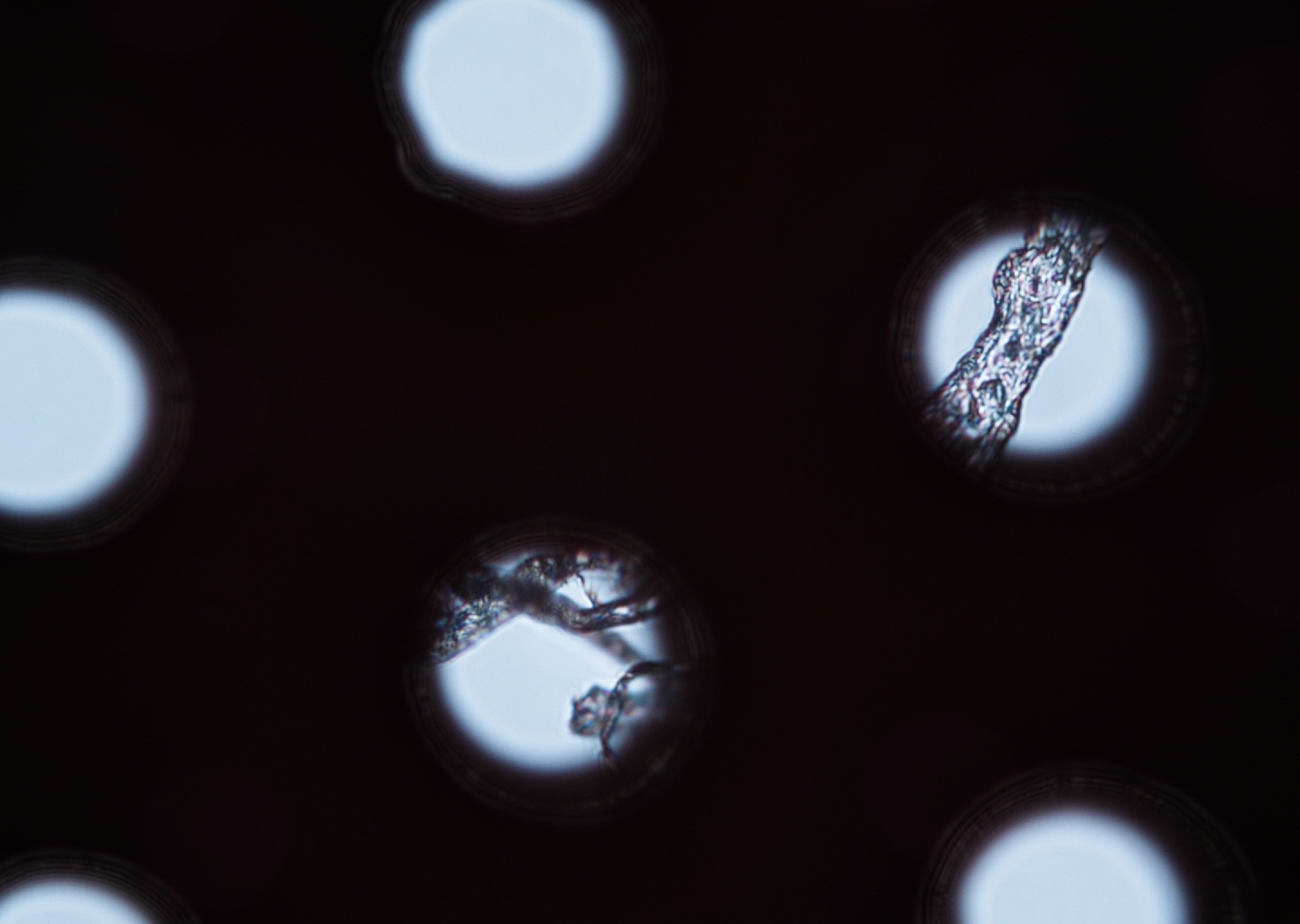}		
			\includegraphics[scale=0.07]{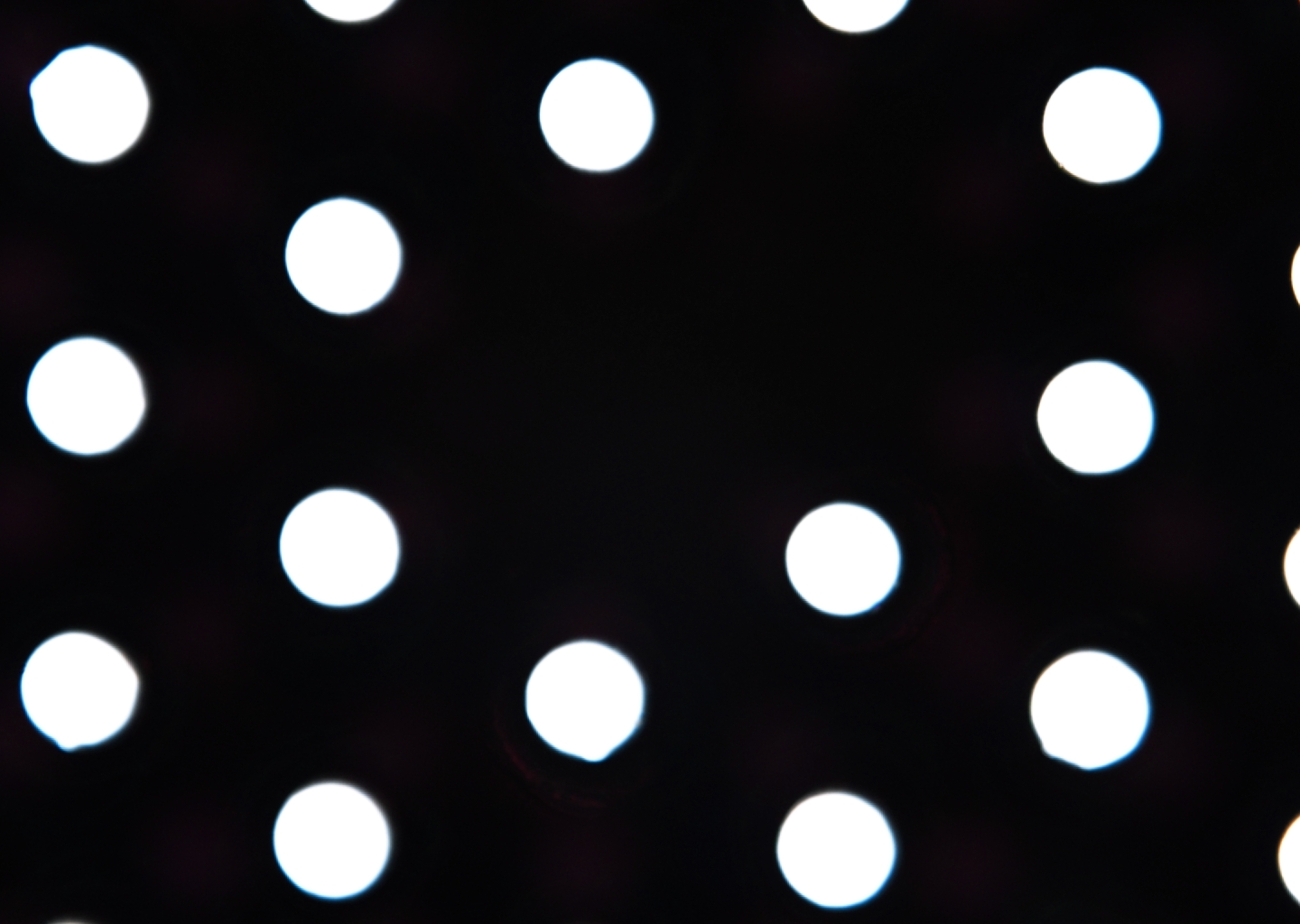}
			\vspace*{-0.3cm}			
			\caption{Microscope setup for scanning the GEM foil (top left). Imperfections in the GEM foil at different magnifications (40X:top right, bottom left; 20X: bottom right)}
			\vspace*{-0.9cm}			
			\label{fig1}
		\end{center}
	\end{figure}
	The microscope setup is shown in Fig.~\ref{fig1}~(top left). The microscope is connected to a PC for taking and storing the image of the object under scanning.
	The visual inspection revealed several imperfections in the foil and they are shown in Fig.~\ref{fig1}. The pitch and diameter of the GEM holes are also measured using the microscope and their distributions are shown in Fig.~\ref{fig2}. The average hole diameter is found to be 69.57~$\underline{+}$~0.09~$\mu$m and the average pitch is found to be 140.20~$\underline{+}$~0.80~$\mu$m. 
	\begin{figure}[htb!]
		\begin{center}
			\vspace*{-0.5cm}
			\includegraphics[scale=0.33]{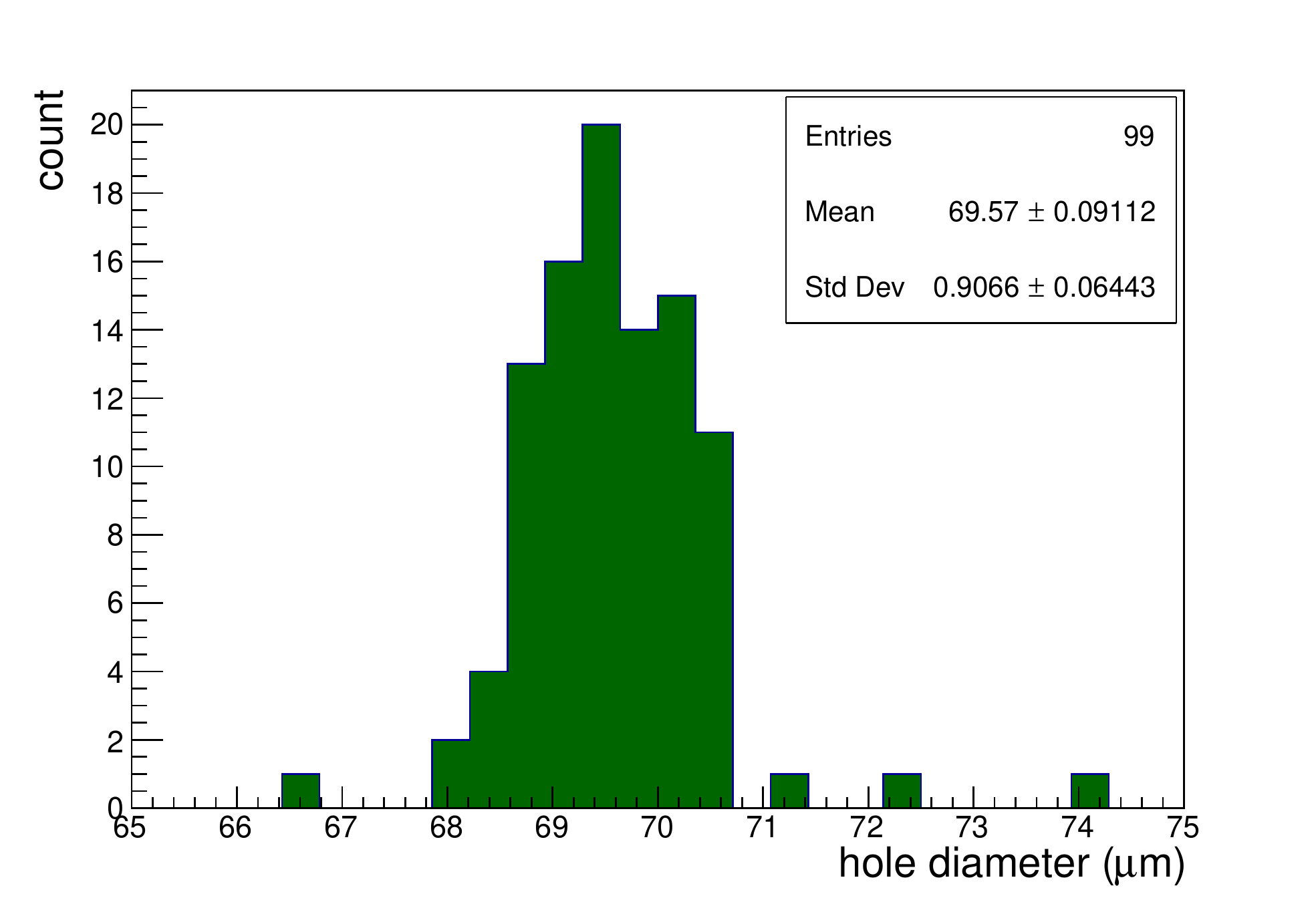}
			\includegraphics[scale=0.33]{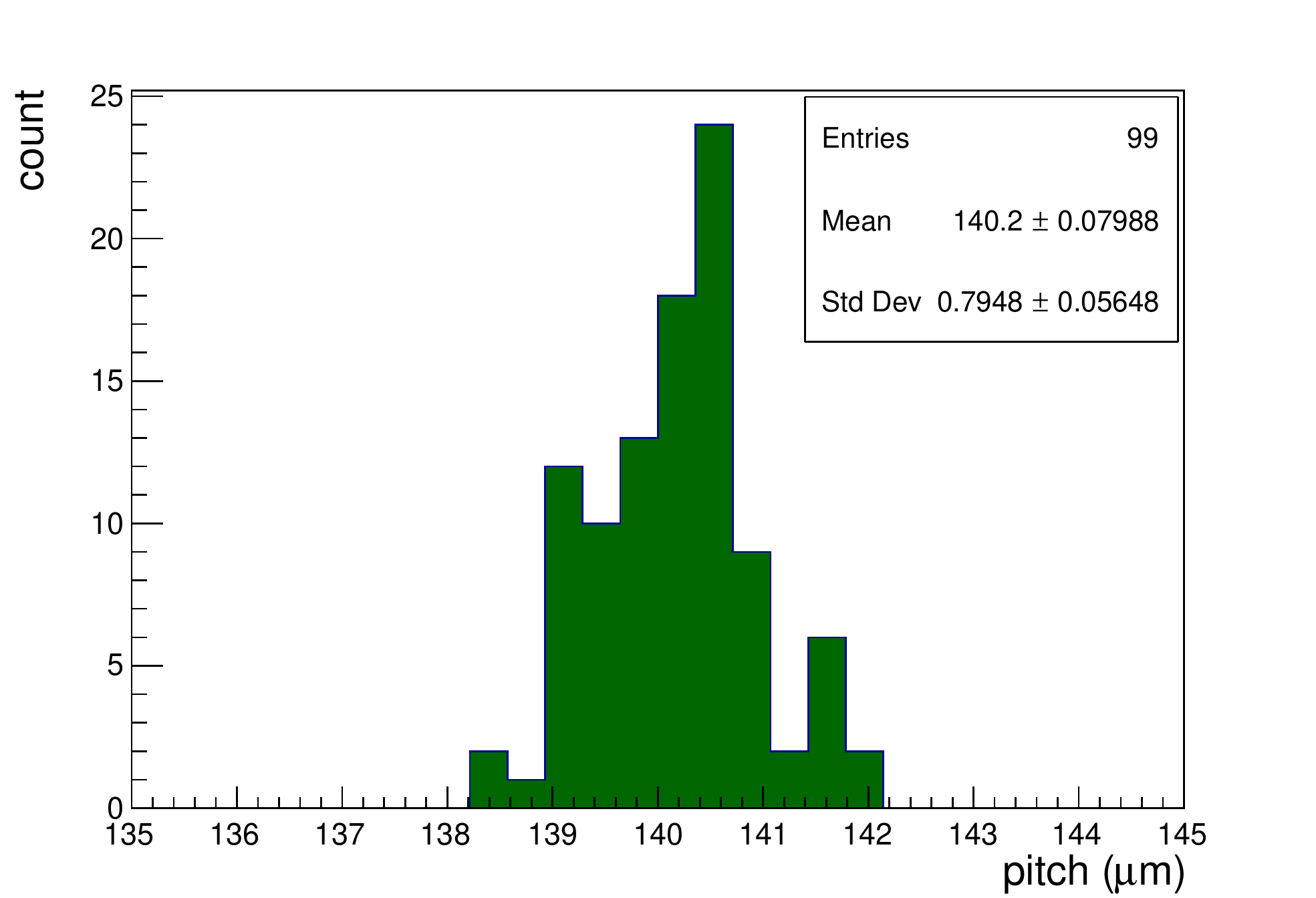}	
			\vspace*{-0.45cm}	
			\caption{Distribution of GEM hole diameter~(top) and pitch~(bottom).}
			\label{fig2}
		\end{center}
	\end{figure}
	
	\vspace*{-0.5cm}
	\section{Methodology for cleaning the GEM foil}\label{section3}
	\vspace*{-0.3cm}
	After doing the visual scanning, the GEM foil is cleaned using millipore water and also using an ultrasonic~($\sim$~20~kHz) bath with millipore water as the medium. 
	
\begin{figure}[htb!]
		\begin{center}
			\vspace*{-0.55cm}	
			\includegraphics[scale=0.08]{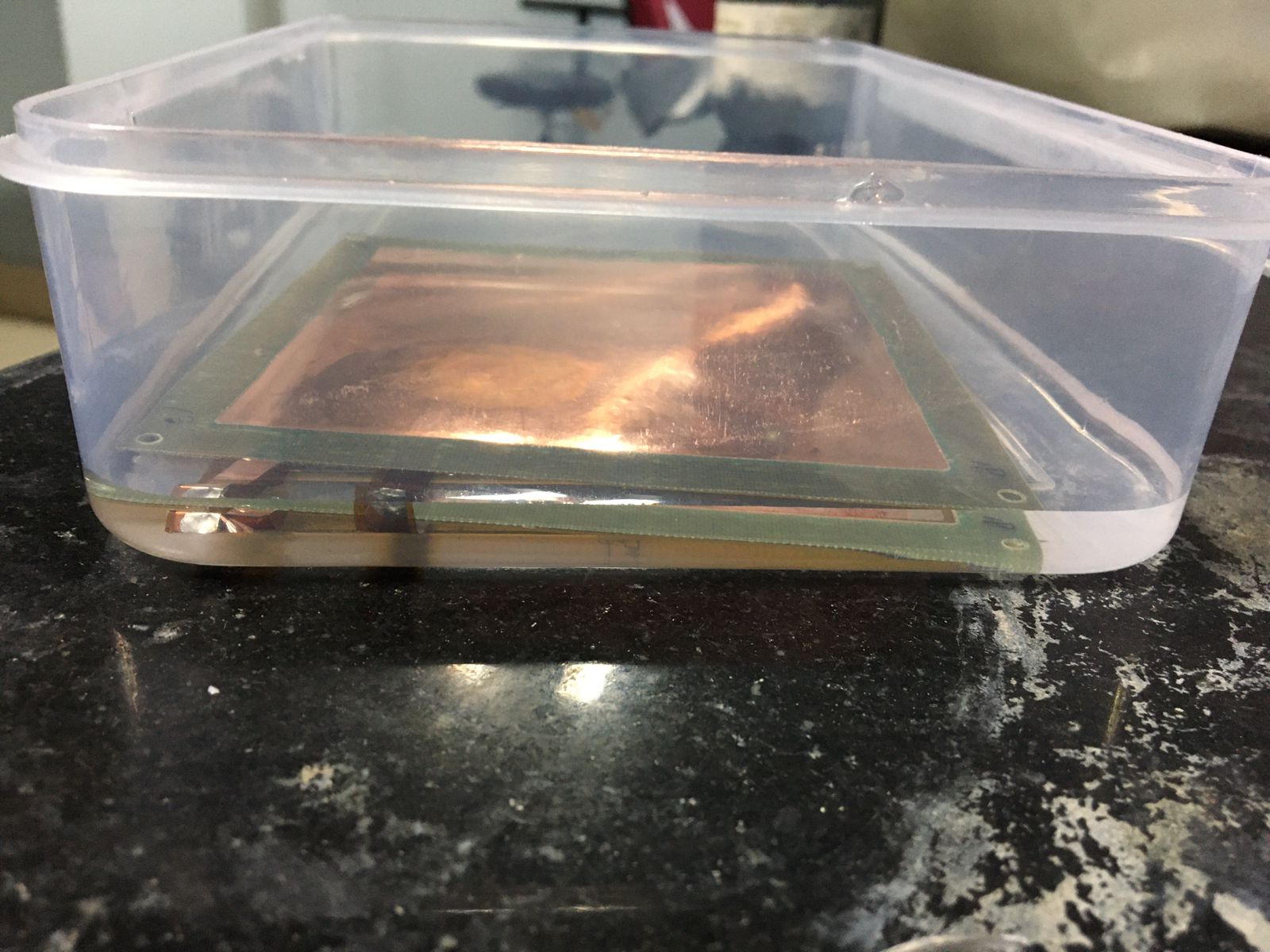}
			\caption{Millipore water bath of the GEM foil.}
			\vspace*{-0.50cm}	
			\label{fig4}
			\vspace{-0.50cm}
		\end{center}
	\end{figure}

	First, the GEM foil is immersed in the millipore water bath for $\sim$~60 minutes and then the foil is removed and kept for drying under continuous hot air flow for $\sim$~30 minutes. After that, the foil resistance is measured and still, it is found that the resistance of the foil is low~($\sim$~1~M$\Omega$). The foil is kept for another 24~hours in a closed box and then again the foil resistance is measured but still, the resistance is found to be $\sim$~1~M$\Omega$. Fig.~\ref{fig4} shows the GEM foil in the millipore water bath.
	
\begin{figure}[htb!]
		\begin{center}
			
			\includegraphics[scale=0.08]{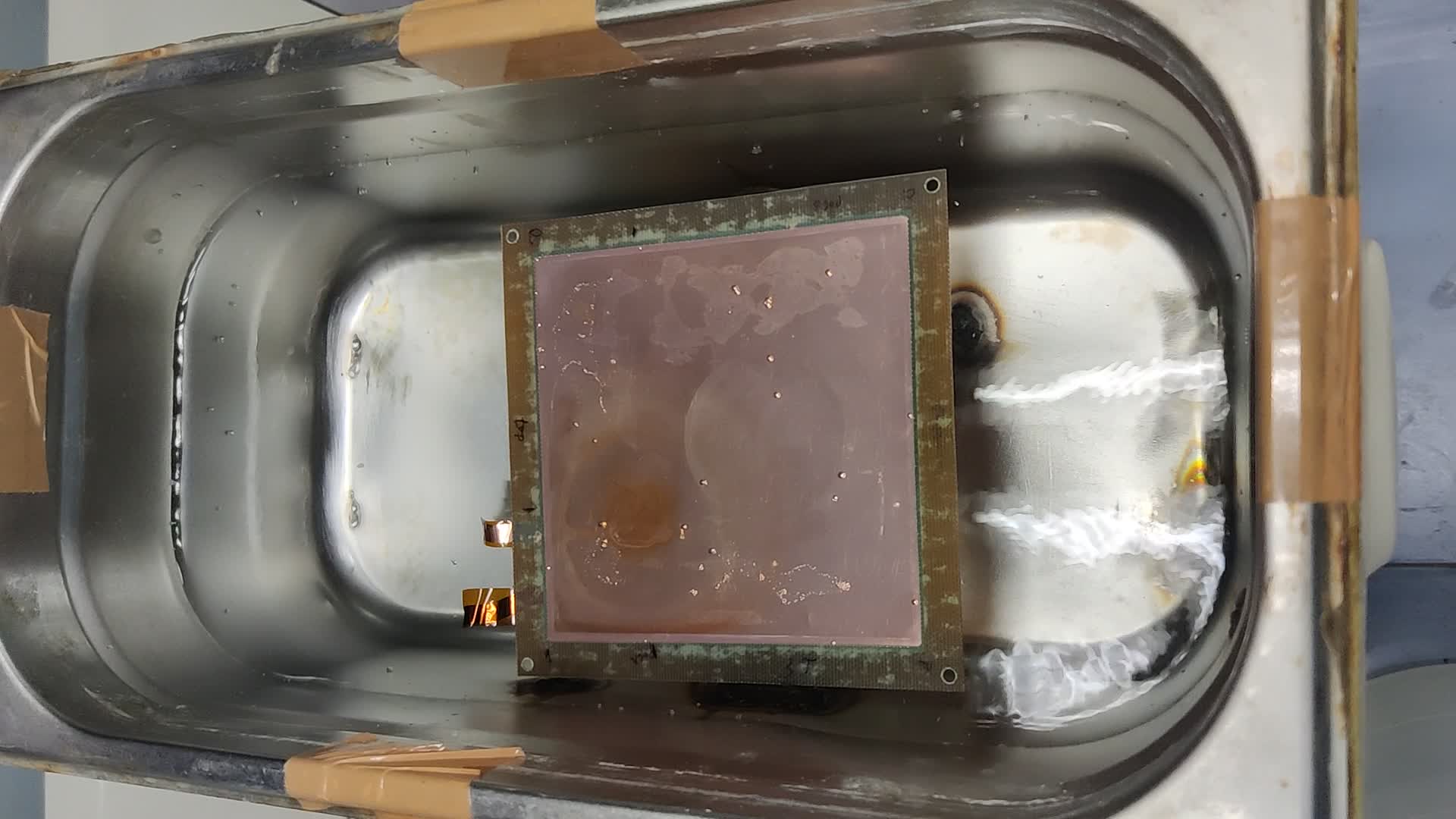}
			\vspace*{-0.3cm}	
			\caption{Ultrasonic bath of the GEM foil with millipore water as the medium.}
			\label{fig5}
			\vspace{-0.40cm}
		\end{center}
	\end{figure}

	After the water bath, the foil is put in the ultrasonic~($\sim$~20~kHz) bath with millipore water as the medium. The foil is kept in the ultrasonic bath for $\sim$~5~minutes. After removing the foil from the ultrasonic bath, the foil is dried for $\sim$~30 minutes under continuous hot air flow. After that, the foil resistance is measured and it is found to be very high, which implies that the short paths are removed. Fig.~\ref{fig5} shows the GEM foil in the ultrasonic bath with millipore water as the medium.
	
	\vspace*{-0.6cm}
	\section{Leakage current measurement of the foil}\label{section4}
\begin{figure}[htb!]
		\begin{center}
			\vspace*{-0.45cm}	
			\includegraphics[scale=0.04, angle=270]{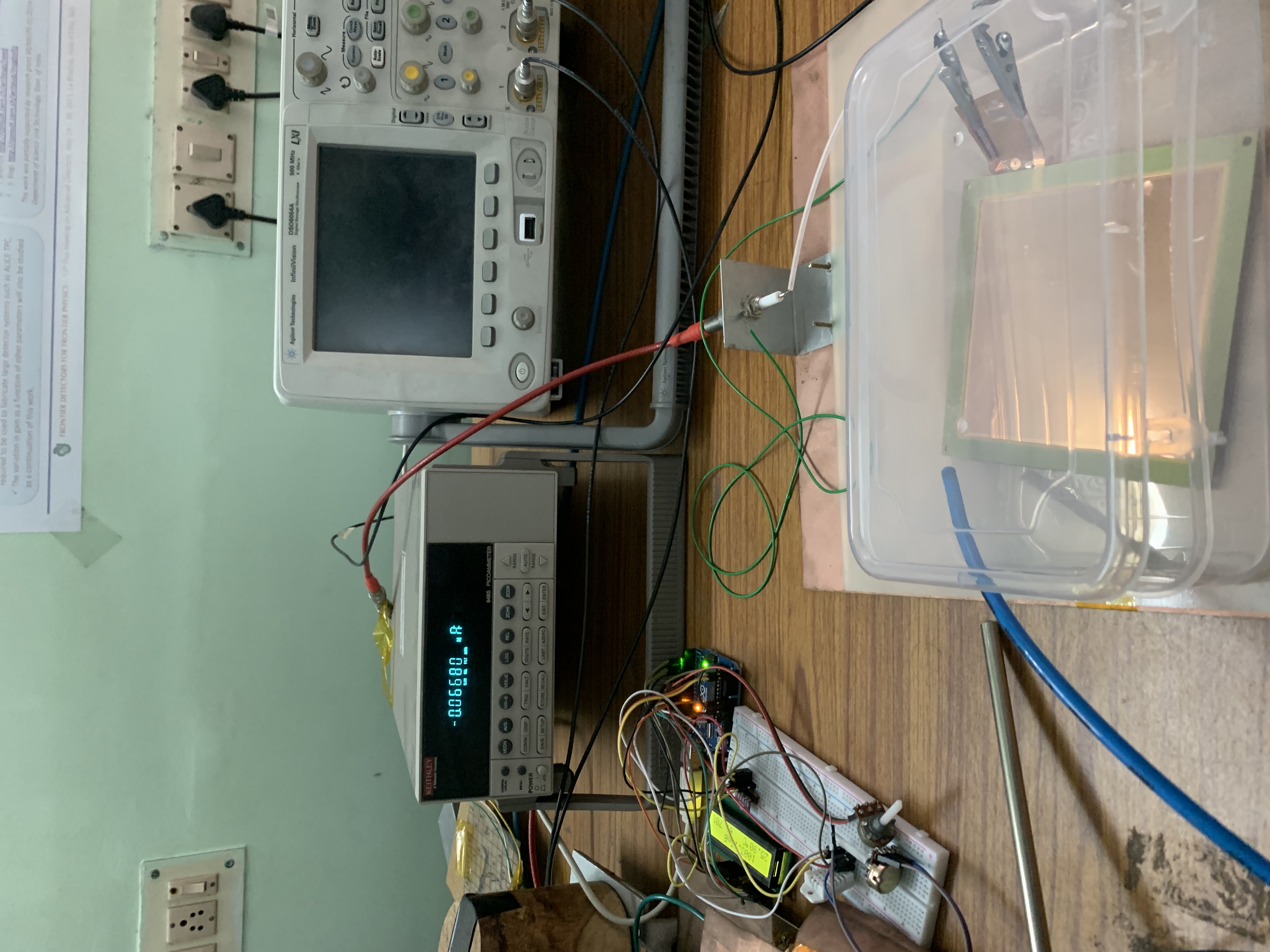}
			\vspace*{-0.3cm}
			\includegraphics[scale=0.35]{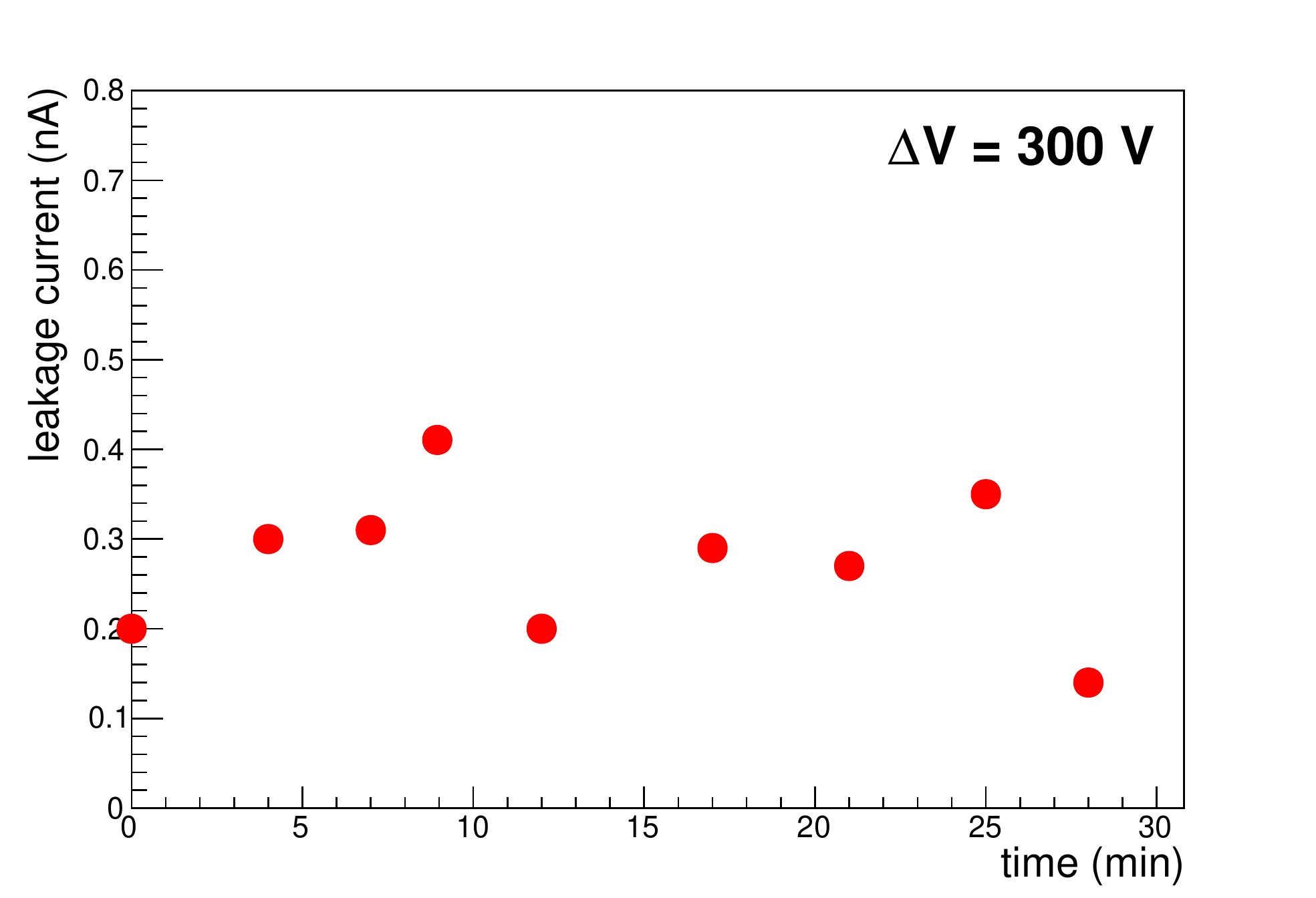}
			\caption{Setup for leakage current measurement of the GEM foil~(top). Leakage current as a function of time (bottom).}
			\label{fig6}
			\vspace{-1.0cm}
		\end{center}
	\end{figure}

	\vspace*{-0.20cm}		
	After cleaning with the ultrasonic bath, the leakage current of the foil is measured under a continuous flow of Ar/CO$_2$ gas mixture. The voltage is applied across the GEM foil by connecting the two leads of the foil to the external High Voltage module. The current is measured using a Keithley picoammeter~(Mod No.: 6485). The setup of the leakage current measurement is shown in Fig.~\ref{fig6}~(top).
	The voltage across the GEM foil is kept at $\Delta$V~$\sim$~300~V and the leakage current of the foil is measured for 30~minutes. The leakage current is found to be $\sim$~0.3~nA at a relative humidity~(RH) of $\sim$~50~\%. 
	The measured leakage current as a function of time is shown in Fig.~\ref{fig6}~(bottom).
	
	\vspace*{-0.50cm}	
	\section{Summary}
	\vspace*{-0.30cm}		
	Visual investigation of a SM GEM foil, showing low resistance~($\sim$~40~k$\Omega$) is carried out manually using an optical microscope. The visual inspection revealed several imperfections in the GEM foil. The GEM foil is cleaned using two different techniques, one by just using the millipore water bath and the other one is with the ultrasonic frequency with millipore water as the medium. After cleaning, the foil resistance is found to be very high. The leakage current of the foil is measured under the continuous flow of the Ar/CO$_2$ gas mixture. The leakage current of the foil is found to be $\sim$~0.3~nA at a $\Delta$V~$\sim$~300~V across the GEM foil and at an RH of $\sim$~50\%. The ultrasonic bath technique is found to be useful to clean the foil showing low resistance, which might be due to the accumulation of impurities in the foil.
	
	\vspace*{-0.60cm}	
	\section{Acknowledgements}
	\vspace*{-0.30cm}	
	
	The authors would like to thank the following persons for their support in the course of this study. Ms. Ruby Biswas, Mr. Dibakar Sarkar, Mr. Pritam	Naskar, Ms. Rudrapriya Das, Mr. Subrata Das, Mr. Mohan Das, Mr. Akash Pandey, Prof. Shubho Chaudhuri, Prof. Anirban Bhunia and Prof. S. K. Ghosh. This work is partially supported by the research grant SR/MF/PS-01/2014-BI from DST, Govt. of India, and the research grant of the CBM-MuCh project from BI-IFCC, DST, Govt. of India. 

\end{document}